\documentclass[twocolumn]{aastex63}

\usepackage[utf8]{inputenc}

\usepackage{amsmath}
\usepackage{amssymb}
\usepackage{gensymb}
\usepackage{cancel}
\usepackage{multirow}
\usepackage{mdframed}

\begin{document}

\title{Simulations of Precessing Jets and the Formation of X-shaped Radio Galaxies}
\author{Chris Nolting}
\affiliation{Los Alamos National Laboratory, Los Alamos, NM, USA}
\affiliation{Department of Physics and Astronomy, College of Charleston, Charleston, SC, USA}
\author{Jay Ball}
\affiliation{Department of Physics and Astronomy, College of Charleston, Charleston, SC, USA}
\author{Tri M. Nguyen}
\affiliation{Department of Physics and Astronomy, College of Charleston, Charleston, SC, USA}

\begin{abstract}

Jet precession is sometimes invoked to explain asymmetries in radio galaxy (RG) jets and ``X/S/Z-shape" radio galaxies, caused by the presence of a binary black hole companion to the source active galactic nucleus (AGN) or by accretion instabilities. We present a series of simulations of radio galaxy jet precession to examine how these sources would evolve over time, including a passive distribution of cosmic ray electrons (CRe) so we can model radio synchrotron emissions and create synthetic radio maps of the sources. We find that a single source viewed from different angles can result in differing RG morphological classifications, confusing physical implications of these classifications. Additionally, the jet trajectories can become unstable due to their own self-interactions and lead to ``reorientation events'' that may look like the effects of external dynamics such as shocks, winds, or cold fronts in the medium. Finally, something akin to an ``Odd Radio Circle" may be observed in the case of viewing the radio remnant of such a precessing source from a line of sight near the precession axis.

\end{abstract}

\section{Introduction}

Radio galaxies consist of an active galactic nucleus (AGN) and a pair of anti-parallel jets of high temperature, low density, supersonic plasma that can propagate to scales larger than the host galaxy. In reality, and especially in denser environments like galaxy clusters, RGs are often distorted, with broken symmetries caused by their interactions with their immediate surroundings. These interactions can be related to motion of the host galaxy relative to a cluster or group center due to an orbit or due to bulk motions within the local medium itself \citep{Begelman79,Jones17}. Another way for these symmetries to be broken is through nonuniform activity in the host AGN, either through variability in the jet properties or changes in its direction. 

One class of RG asymmetries that is common and being studied more frequently is the so called ``X-shaped'' RG \citep{Leahy92,Bhukta22}. These RGs have two pairs of radio lobes that are misaligned from each other. Sometimes one pair is designated as the ``main lobes'' and the other pair as ``wings'' or ``side-lobes,'' with the designation based on the detection of active radio jets or radio hot spots in the main lobes, or from surface brightness or spectral aging considerations.

There have been multiple suggested theories for the formation of such structures. Some suggest that RG wings can form through the backflow of radio plasma that rebounds laterally off the hot gaseous halo of the host galaxy \citep{Leahy84,Cotton20,Ignesti20}. Others invoke a ``spin-flip'' in the AGN due to a merger event between a binary supermassive black hole (BSBH) system and a sudden reorientation of the spin axis of the active nucleus \citep{Zier01,GopalKrishna12}. It may also be possible for both nuclei in a BSBH system to be accreting, in which case two sets of jets may exist. These may both be active during the same period or undergoing some cadence of activity depending on their respective accretion histories, creating multiple sets of lobes \citep{GopalKrishna12}.

Lastly, and most relevant to what we will discuss here, is the idea that the jet axis of the RG may change or precess over time. This precession may lead to the formation of X-shaped RGs if the precession angle is large so that the current axis is well separated from the side lobes. Smaller angles may lead to ``S-shaped'' or ``Z-shaped'' RGs, which are designations sometimes given to RGs that show curvature or sharp turns in their jets or lobes \citep{Riley72}.

There are many examples of S-shaped radio sources that could be fit by a projected helical precessing jet such as 4C35.06 in Abell 407 \citep{Biju14}, J1328+2752 \citep{Nandi21}, or Hydra A \citep{Taylor90}. On smaller scales, SS433 is a striking example of a precessing X-ray binary jet system \citep{Abell79, Monceau-Baroux14}. The precession of AGN jets on large scales has been studied extensively as a method of creating X-shaped RGs \citep{Ekers78,Rubinur17}. In particular, (magneto)hydrodynamical simulations have proven useful in characterizing the evolution and morphology of such jets \citep{Smith19,Horton20,Giri22}. 

Precession of an RG jet may occur for a number of reasons. Again invoking a BSBH system, if either nucleus is substantially accreting and hosts an active radio jet, then the jet may precess around the orbital axis of the BSBH system, assuming there is a misalignment between the orientation of the binary orbit and the black hole spin axis \citep{Begelman80}. Alternatively, precession may be related to accretion instabilities such as Lense-Thirring or the related Bardeen Petterson effect \citep{Bardeen75,Nandi21}. These instabilities may cause the precession of the accretion disk if its angular momentum is misaligned with the angular momentum of the central spinning black hole. If the disk orientation controls the jet axis, then this would induce a precession in the jet axis. This precession mechanism is seen in some general relativistic accretion disk simulations \citep[e.g.,][]{Liska18}.

Regardless of the mechanism for the precession of the jet, we present the results of simulations in which we assume jets do precess, and discuss the consequences of that precession for the radio observables from these simulated RGs. One of our main goals is to use these simulations to invert this process and be better able to understand the dynamical state of observed radio jets that appear to be precessing. 

 The remainder of this paper is organized as follows: Section \ref{sec:precessioncartoon} outlines the geometry of the interaction we are examining and the underlying physics. Section \ref{sec:methods} describes the simulations, including methods and specific parameters for each simulation. We describe the results of the simulations in section \ref{sec:Discussion} and summarize the major findings in section \ref{sec:Summary}.

\section{Jet Physics}
\label{sec:precessioncartoon}

\subsection{Jet Precession}

\begin{figure}
    \centering
    \includegraphics[width=0.5\textwidth]{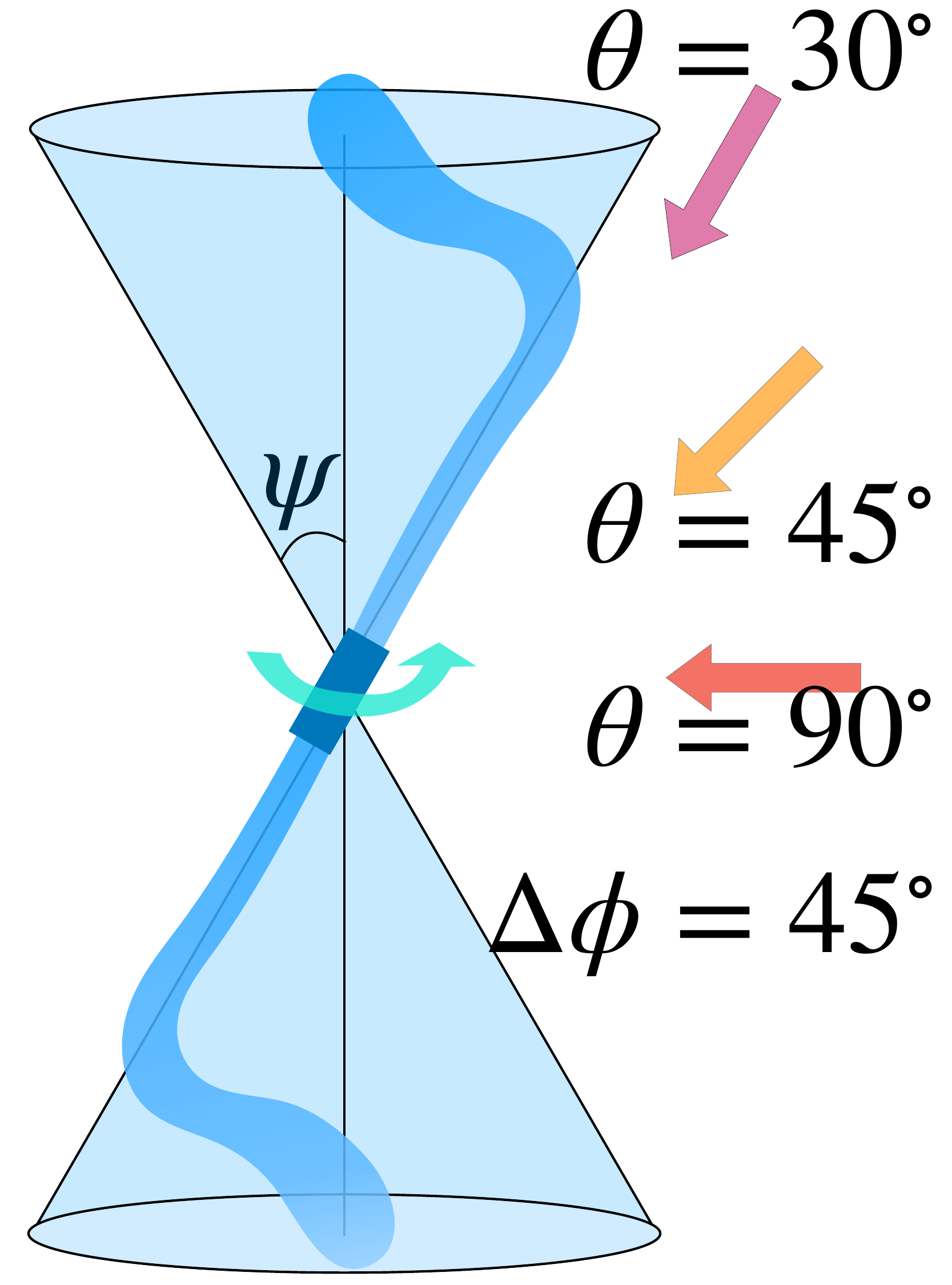}
    \caption{Diagram of the jet precession. The precession angle $\psi$ is the angle between the jet axis and the precession axis, and the diagram uses a fiducial value of $\psi = 30\degree$. The polar viewing angle, $\theta$, is the angle between the precession axis and the viewing angle, with angles used in Figure \ref{fig:jpr03-angles} shown. Not shown are the azimuthal viewing angles which are rotated around the precession axis in increments of $\Delta\phi = 45^\circ$. A cyan arrow indicates the direction that the jet axis precesses.
    \label{fig:precession}}
\end{figure}

The basic geometry of the type of interactions we explore in this paper are illustrated in Figure \ref{fig:precession}. The jet is launched from a cylindrical region in which the cylinder axis is tilted by $\psi$ degrees from the z-axis of the simulation, which we will refer to as the precession angle. The other angles in Figure \ref{fig:precession} refer to the viewing angles used in the synthetic radio observations we present in Section \ref{sec:Discussion}. $\theta$ is the spherical coordinate polar angle, also measured from the z-axis. Observations are taken at four $\theta$ angles: $90\degree$, $45\degree$, $30\degree$, and $0\degree$. $\phi$ is the azimuthal angle, and observations are taken at four $\phi$ angles from $0\degree \le \phi \le 135\degree$ in $\Delta \phi = 45 \degree$ intervals. 

The jet will precess through a full 2$\pi$ radians within one precession period, $\tau$. This can also be expressed as an angular velocity of precession, $\omega_{prec} = 2\pi/\tau$.

\subsection{Jet Propagation \& Bending}
\label{subsec:propgation}

As the jet precesses, it will bend due to its interaction with the medium. The character of this bending will be determined by a combination of the momentum in the jet and the properties of the precession of its launching angle.

The forward propagation of a jet can be described by a momentum balance in the frame of the head of the jet. As derived in \cite{Jones17}, we find that in the absence of relative motion between the surrounding medium and the jet source, the propagation of the head of the jet will follow:
\begin{equation}\label{eq:momentum1}
    M_h \approx M_j \sqrt{\frac{A_j}{A_a}} \sqrt{\frac{P_j}{P_a}},
\end{equation}
where $M_h = v_h/c_{s,a}$ is the Mach number of the head of the jet propagating at velocity $v_h$ relative to the ambient medium that has a sound speed of $c_{s,a} = \sqrt{\gamma P_a/\rho_a}$. $M_j= v_j/c_{s,j}$ is the Mach number of the material within the jet that has velocity $v_j$ and has an internal sound speed of $c_{s,j} = \sqrt{\gamma P_j/\rho_j}$. $P_{(j/a)}$ and $\rho_{(j/a)}$ are the thermal pressure and density of the (jet/ambient) plasma, respectively, and we use an adiabatic index of $\gamma = 5/3$. $\sqrt{A_j/A_h}$ is a ratio of ``effective areas" over which the momentum balance is taken for the jet and ambient plasma. In previous simulations of steady, non-precessing jets, we have empirically found $\sqrt{A_j/A_h} \sim 1/2$ \citep{Jones17}. This ratio departs from unity mainly due to the existence of backflow of jet plasma away from the region of the jet head. In the case where the jet is in pressure equilibrium with its surroundings, i.e. $P_j = P_a$, Equation \ref{eq:momentum1} can be expressed as:
\begin{equation}\label{eq:momentum2}
    v_h \approx v_j \sqrt{\frac{A_j}{A_h}} \sqrt{\frac{\rho_j}{\rho_a}},
\end{equation}
which is a useful metric when the jet head crosses a density discontinuity. 

When there is some relative motion between the source of the jet and its environment, the propagation will be affected. If that motion has a component transverse to the jet propagation direction, then the jet may be deflected or bent by the ram pressure it experiences. The classic ``cartoon" formula for the distance, $\ell_b$, over which the jet is bent backwards by a ``wind" is \citep{Begelman79,Jones17}:
\begin{equation}\label{eq:bending1}
    \ell_b \sim 2r_j \frac{\rho_jv_j^2}{\rho_av_a^2},
\end{equation}
where $r_j$ is the cross-sectional radius of the jet and $v_a$ is the ambient velocity that provides the ram pressure to bend the jet. Equation \ref{eq:bending1} assumes a constant $v_a$, but if we assume this ``wind" comes from the precession of the jet through a uniform medium at rest, then the relevant ambient velocity is defined by the angular velocity of the precession of the jet source. $v_a(y) = \omega_{prec} \cdot y \sin{\psi} = 2\pi y \sin{\psi}/\tau$, where $y$ is the distance along the jet direction from the jet source to the point of interest. Since the ambient velocity varies along the jet, we need to revisit the arguments that lead to equation \ref{eq:bending1}. Usually, this begins by assuming that the ram pressure acts to redirect the jet, without compressing or accelerating the jet plasma. This leads to an incompressible Euler equation:
\begin{equation}\label{eq:eulerBending}
 \frac{D v_j}{D t} = \cancelto{0}{\frac{\partial v_j}{\partial t}} +  (v_j\cdot \nabla)v_j = - \frac{1}{\rho_j} \frac{\partial P}{\partial x} \approx - \frac{1}{\rho_j}  \frac{\rho_a v_a^2}{2 r_j},
\end{equation}
where we assume that the jet bending leads to a steady state ($\partial v_j/\partial t = 0$) and that the ram pressure, $ P = \rho_a v_a^2$, acts across the diameter of the jet, $2r_j$. If we look at the component of the jet velocity perpendicular to the jet launching axis ($v_{j,\perp}$) and integrate along the jet length until the jet has been bent into the direction of the ``wind,'' i.e. $v_{j,\perp} = v_j$:
\begin{equation}\label{eq:bendingIntegral}
    \int_{0}^{v_j} dv = \frac{1}{2r_j}\frac{\rho_a}{\rho_j v_j} \left(\frac{2\pi \sin{\psi}}{\tau}\right)^2 \int_{0}^{\ell_b} y^2 dy,
\end{equation}
\begin{equation}\label{eq:bending2}
    \ell_b \sim \left(r_j\frac{3\rho_jv_j^2\tau^2}{2\pi^2 \rho_a \sin^2{\psi}}\right)^{1/3},
\end{equation}
showing that the bending length of the jet will be a function of the jet properties, the precession angle and period, and the density of the medium. 

Now, using Equation \ref{eq:momentum2}, we can compare this length to the distance that the head of the jet would travel before the precession has significantly changed its launching direction (ignoring for now how the bending affects this propagation). For this duration, we will take the fraction of the total period that it takes the core of the jet to transverse one jet diameter. This will depend on the shape and size of the jet launching cylinder:
\begin{equation}\label{eq:dt}
    \Delta t = \frac{2r_j}{2\pi \Delta y \sin{\psi}} P,
\end{equation}
where $\Delta y$ is the half length of the cylinder (from the center to one end). The ratio $\ell_b/(v_h \Delta t)$ is then a good measure of whether or not the jet will create a well defined curved jet, or if the precession will be too extreme and cause the jet to break up or be unable to propagate to large scales.
\section{Numerical Methods}
\label{sec:methods}

The simulations reported here used the Eulerian WOMBAT ideal 3D MHD code (see, e.g. \cite{Mendygral12,Nolting19a}) on a uniform, Cartesian grid employing an adiabatic equation of state with $\gamma = 5/3$. The simulations utilized the 2$^{nd}$ order TVD algorithm with constrained transport (CT) magnetic field evolution as in \cite{Ryu98}. Specific simulation setups are introduced in \S \ref{subsec:Setup} and listed in Table \ref{tab:parameters}. 

Along with the fluid, we track a population of passive cosmic ray electrons (CRe) to allow for the calculation of a synchrotron emissivity anywhere within the simulation volume\footnote{Except for a negligible ICM population included to avoid numerical singularities in the CRe transport algorithm, all CRe were injected onto the computational domain via the jet launch cylinder.}. The CRe momentum distribution, $f(p)$, was tracked using the conservative, Eulerian ``coarse grained momentum volume transport''  CGMV algorithm in \cite{JonesKang05}. $f(p)$ spanned the range $10 \la p/(m_e c)\approx \Gamma_e\la 1.7\times 10^5$ (so, energies 5 MeV $\la E_{CRe} \approx \Gamma_e m_e c^2 \la$ 90 GeV) with uniform logarithmic momentum bins, $1\le k\le 8$. Inside a given momentum bin, $k$, $f(p) \propto p^{-q_{k}}$, with $q_k$ being bin dependent and evolving in time and space. $\Gamma_e$ here represents CRe Lorentz factors. 

At the jet cylinder, the CRe momentum distribution had a power law form with index $q = q_0 = 4.2$ across the full momentum range simulated. This was chosen because this translates to a radio synchrotron spectral index of $\alpha = \alpha_0 = 0.6$ ($I_{\nu} \propto \nu^{-\alpha})$, which is a good match for many RGs near their sources. The synchrotron emission and spectra reported here are calculated numerically using $f(p)$ over the full momentum range specified above using the standard synchrotron emission kernel for isotropic electrons in a local magnetic field \citep[e.g.,][]{BlumenthalGould70B}. In our analysis we calculated synthetic synchrotron emission at frequencies $300$ MHz $\leq \nu$ $\leq 600$MHz. This emission, as it turns out, comes predominantly from regions with magnetic field strengths $\sim$ a few $  \mu$G, so mostly reflect CRe energies $\ga$ 1 GeV ($\Gamma_e \sim 10^3 $--$ 10^4$) (well inside our distribution).

We included adiabatic, as well as radiative (synchrotron and inverse Compton) CRe energy changes outside of shocks, along with test-particle diffusive shock (re)acceleration (DSA) at any shocks encountered. We did not include $2^{nd}$ order turbulent CRe reacceleration or CRe energy losses from Coulomb collisions with ambient plasma. The former depends on uncertain kinetic scale turbulence behaviors beyond the scope of this study, while the latter is most relevant for CRe with energies well below those responsible for the radio synchrotron emission computed in this work \citep{Sarazin99}. CRe radiative losses combine synchrotron with inverse Compton (iC) scattered CMB radiation. The simulations reported here assumed a low redshift, $z = 0.02$. The resulting radiative lifetime can be written
\begin{equation}\label{eq:lifetime}
\tau_{rad} \approx 215 \frac{1}{\Gamma_{e4}\left[1+ B_{3.4}^2\right]}~\rm{Myr},
\end{equation}
where $\Gamma_{e4} = \Gamma_e/10^4$ and $B_{3.4} = B/(3.4\mu\rm{G})$.  The first term in the denominator on the RHS reflects inverse Compton (iC) losses at z = 0.02, while the second represents synchrotron losses. Thus, we can see that for $\Gamma_e \sim 10^4$, of primary interest for the radio emission in this work, $\tau_{rad} \sim 200$ Myr, and that iC losses are predominant.

DSA of the CRe was implemented at shock passage by setting $q_{k,out} = \min(q_{k,in},3\sigma /(\sigma - 1))$ immediately post-shock, where $\sigma$ is the code-evaluated compression ratio of the shock. This simple treatment is appropriate in the CRe energy range covered, since likely DSA acceleration times to those energies are much less than a typical time step in the simulations ($\ga 10^4$ yr). Since our CRe have no dynamical impact, we treat the total CRe number density, $n_{CRe}$, as arbitrary. Consequently, while we compute meaningful synchrotron brightness and spectral distributions from our simulations, synchrotron intensity normalizations are arbitrary. 

\subsection{Simulation Setup}
\label{subsec:Setup}

In this work we present a suite of simulations that explore the parameter space of jet precession angles and precession periods. Each simulation consists of a pair of oppositely oriented jets, each with constant density, pressure, and velocity being injected into a medium that is initially homogeneous and unmagnetized. This medium is an oversimplification of the intracluster medium (ICM), but it is useful for easier interpretation of the observable signatures of the precession that we wish to study. The dynamics of the simulations we present here are scale free, however a scale is set when introducing CRe radiative timescales (e.g., Equation \ref{eq:lifetime}). Because of this, we will quote the scale of the system that is consistent with the chosen radiation timescales.

Inside the simulation volume, a cylindrical region was updated at each time step to impose a set of jet properties. The jets in this paper all had an initial density of $\rho_j = 1.1 \times 10^{-29}$ g cm$^{-3}$, a velocity of $ v_j \approx 0.1$c $ = 3.0\times 10^{9} $ cm s$^{-1}$ at the end of the cylinder, and a pressure in equilibrium with their surroundings $P_j = 8.8$ dyne cm$^{-2}$. This jet launching cylinder had a radius of $r_j = 3$kpc and length $l_j = 12$kpc. The jet cylinder was surrounded by a 2 zone coaxial collar, within which the state transitioned from the jet properties to the local ambient conditions. Inside, the jet density and pressure were kept constant, and the velocity was ramped up along the cylinder's length from its midpoint (where the velocity reverses). A toroidal magnetic field was also maintained withing the jet cylinder, with $B_{\phi} = B_j (r/r_j)\hat{\phi}$. In these jets, $\beta_p = 8\pi P_j/B_j^2 = 75$, giving the magnitude of the field at launch to be $B_j \approx 0.54\mu$G. Due to this relatively high $\beta_p$, the fields are initially dynamically sub-dominant to the gas pressure, which remains true throughout the simulation duration in all cases. 

To create the precession of the jet axis, the jet is initialized $\psi$ degrees away from the precession axis, defining the precession angle. Every time step, the jet cylinder is rotated azimuthally around the precession axis by $\Delta \phi = 2\pi \Delta t / \tau$, where $\Delta t$ is the length of the time step and $\tau$ is the jet precession period.

Each simulation had a uniform initial density of $\rho_{ICM} = 2.17\times10^{-28}$ g cm$^{-3}$, and pressure of $P_{ICM} = 8.8\times 10^{-13}$ dynes cm$^{-2}$ and was initially unmagnitized and at rest with respect to the jet source. This gave the medium an initial temperature of 2.5keV.

Table \ref{tab:parameters} gives a list of parameters for the 12 simulations, including domain size, precession period, and precession angle. Three precession periods and four precession angles were included in this study. Simulation names include information about the precession period (number following ``P" in the name) and the precession angle (number following ``A" in the name) for ease of understanding when referring to individual simulations.


\begin{table}
    \centering
    \begin{tabular}{c|c|c|c|c|c}
       Simulation & \multirow{2}{*}{Nx} & \multirow{2}{*}{Ny} & \multirow{2}{*}{Nz} & Precession & Precession\\
       Name & \hfill & \hfill & \hfill & Period (Myr) & Angle \\\hline
       
         \textbf{P3A10} & 288 & 288 & 624 & 3.2 & 10\degree\\ 
         \textbf{P3A20} & 288 & 288 & 624 & 3.2 & 20\degree\\ 
         \textbf{P3A30} & 288 & 288 & 624 & 3.2 & 30\degree\\ 
         \textbf{P3A45} & 360 & 600 & 600 & 3.2 & 45\degree\\ 
         \textbf{P32A10} & 288 & 288 & 624 & 31.7 & 10\degree\\ 
         \textbf{P32A20} & 288 & 288 & 624 & 31.7 & 20\degree\\ 
         \textbf{P32A30} & 288 & 288 & 448 & 31.7 & 30\degree\\ 
         \textbf{P32A45} & 360 & 600 & 600 & 31.7 & 45\degree\\ 
         
          \textbf{P95A10} & 288 & 288 & 1040 & 95.1 & 10\degree\\ 
         \textbf{P95A20} & 288 & 288 & 624 & 95.1 & 20\degree\\ 
         \textbf{P95A30} & 288 & 288 & 448 & 95.1 & 30\degree\\ 
         \textbf{P95A45} & 360 & 600 & 600 & 95.1 & 45\degree\\ 
        
    \end{tabular}
    \caption{Names, domain size, and precession parameters for each simulation.}
    \label{tab:parameters}
\end{table}

\subsection{Synthetic Observations}

The radio images we report here are calculated from the CRe momentum distribution $f(p)$ multiplied by the appropriate synchrotron emissivity functions integrated across the full momentum range, as specified above. This emissivity takes into account the strength and orientation of the local magnetic field, as well as the history of the CRe energy changes (radiative, adiabatic, \& DSA). The emissivity is calculated as \citep{Ginzburg65,Longair11}:

\begin{subequations}
\small
 \begin{align}
j_\perp(\nu) & =  \sqrt{\frac{\nu \nu_{B\perp}}{8}} \frac{e^2}{c}  \int_{x_1}^{x_2} \frac{n(x) [F(x) + G(x)]}{x^{3/2}} \mathrm{d}x \label{eq:jperp} ~,\\
j_\parallel(\nu) &  =  \sqrt{\frac{\nu \nu_{B\perp}}{8}} \frac{e^2}{c}  \int_{x_1}^{x_2} \frac{n(x) [F(x) - G(x)]}{x^{3/2}} \mathrm{d}x \label{eq:jpar} ~,\\
j(\nu) & = j_\perp(\nu) + j_\parallel(\nu) =  \sqrt{\frac{\nu \nu_{B\perp}}{2}} \frac{e^2}{c}  \int_{x_1}^{x_2} \frac{n(x) F(x)}{x^{3/2}} \mathrm{d}x \label{eq:jnu} ~,\\
F(x) & = x \int_x^\infty K_{5/3}(z) \mathrm{d}z \approx 1.78 \nonumber \\ 
&\times \left(\frac{x}{1 - 0.4 \exp{(-5x)}}\right)^{1/3}\exp{(-x)} \label{eq:Ffit} ~,\\
G(x) & = x K_{2/3}(x) \approx 1.56613 \times \frac{x^{1/3}}{\exp{(x) + 0.427687}} ~,\label{eq:Gfit}
 \end{align}
\end{subequations}
where $F(x)$ and $G(x)$ are functions that describe the synchrotron spectrum of a single CRe, $K_{5/3}$ and $K_{2/3}$ are modified Bessel functions of the second kind, $j_\perp(\nu)$ and $j_\parallel(\nu)$ are the two orthogonal polarizations of the emissivity, $\nu$ is the frequency of the radiation, $\nu_{B\perp} = eB_\perp/2\pi m_e c$ is the electron gyrofrequency, and $e$ is the elementary charge. In this case, $B_\perp$ is the local magnetic field projected onto the plane of the sky.
The integration variable $x = (2\nu/3\nu_{B\perp}\gamma^2)$, with $\gamma$ being the electron Lorentz factor. The integration limit $x_1$[$x_2$] corresponds to the high[low] energy (high[low] $\gamma$) end of the integration range, covering our available range $10 < p/(m_ec) < 1.7 \times 10^5$. A change of variables from integrating over $\gamma$ to integrating over $x$ introduces the factor of $x^{-3/2}$ in the integrand and a coefficient of $(-1/2)\sqrt{2\nu/3\nu_{B\perp}}$. We also present these equations in Gaussian units, rather than the SI units used by \cite{Longair11}.
Equations (\ref{eq:Ffit}) and (\ref{eq:Gfit}) include our own approximations to the synchrotron functions used in our calculations. These approximations are based on those given in \cite{RybickiLightman86}, but combine the small $x$ and large $x$ approximations to produce a single fit that is accurate to a few percent \citep{NoltingThesis}. 

Using equations (\ref{eq:jperp}) - (\ref{eq:Gfit}) we calculate the total synchrotron emissivity as well as polarized emissivities. Then, we perform radiative transfer integration along a defined line of sight to create images of Stokes I, Q, or U. 

A radio spectral index is calculated from any two radio maps at different frequencies with the radio spectral index, $\alpha$ (e.g., $f\propto \nu^\alpha$), at each pixel being
\begin{equation}\label{eq:index}
    \alpha = \frac{\log_{10}(I(\nu_2) - \log_{10}(I(\nu_1)))}{\log_{10}(\nu_2) - \log_{10}(\nu_1)}.
\end{equation}
In this work, the two frequencies used to generate spectral index maps were $\nu_2 = 600$ MHz and $\nu_1 = 300$ MHz. Radio spectral index maps are shown weighted by the radio intensity at 300MHz. This was done by applying an alpha transparency filter to the spectral index map according to the weighted brightness. The weight was calculated as:
\begin{equation}\label{eq:weights}
    w_{i,j} = \sqrt{\min(I_{300\text{MHz},i,j},I_{max})}
\end{equation}
where $I_{max}$ was the $90^{th}$ percentile brightest pixel in the image (or reference time/frame in images or animations with multiple frames, see figure captions).

Lastly, all radio images presented here are convolved with a gaussian convolution kernel. The resolution for most images is 2.355 arcsec Full Width Half Max (FWHM), or a gaussian standard deviation of 1 arcsecond, while one figure is convolved to a lower resolution of 11.775 arcsec FWHM, or 5 arcseconds gaussian standard deviation. 

\section{Discussion}
\label{sec:Discussion}

\begin{figure*}
    \centering
    \includegraphics[width=0.85\textwidth]{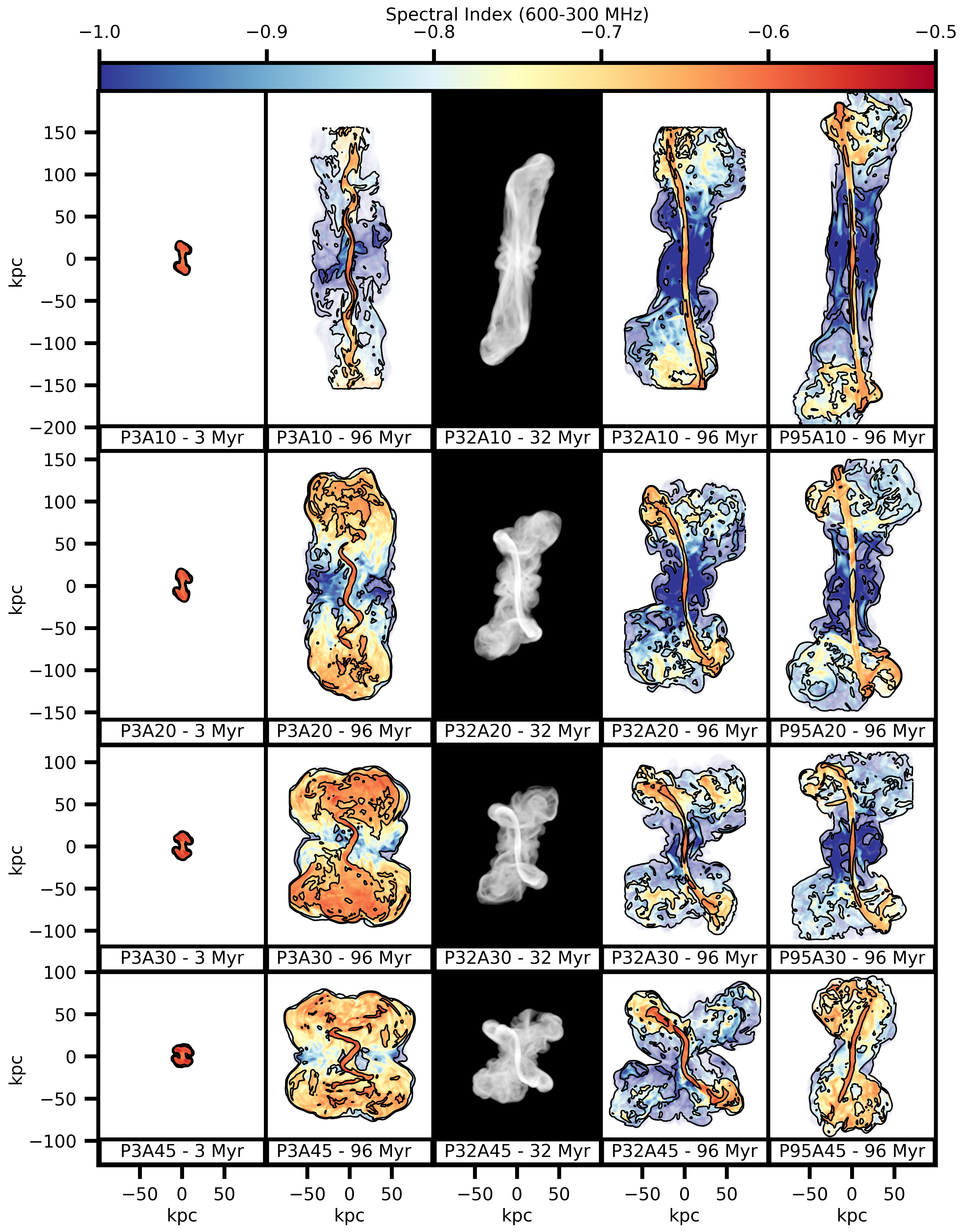}
    \caption{Columns 1, 2, 4, and 5: Radio spectral index (600-300MHz, 
    weighted by 300MHz intensity with $I_{max}$ based on the bottom right frame, simulation P95A45)
    with 300 MHz radio brightness contours (levels = [1, 10, 100]$\mu$Jy beam$^{-1}$). Column 3, which varied least in spectral index ($\sim 0.6$ in the jets, $\sim 0.75$ in the lobes), shows instead the radio brightness at 300MHz. Radio images are convolved to 2.355'' FWHM resolution. Each row shows simulations with different precession angles (row 1: 10$\degree$, 2: 20$\degree$, 3: 30$\degree$, 4: 45$\degree$). Columns show simulations with differing precession periods, or the same simulations at different times (period for column 1 \& 2: 3.2 Myr, 3 \& 4: 31.7 Myr, 5: 95.1 Myr) (column 1, 3, \& 5 are shown at approximately 1 precession period; columns 2, 4, \& 5 are shown at 96 Myr simulation time). }
    \label{fig:comparison-mosaic}
\end{figure*}

\subsection{Effects of Precession Period and Precession Angle}

To study the effects of precession period and precession angle, ran a simulation study varying these parameters. The resulting morphological and radio spectral properties from these simulations are presented in figure \ref{fig:comparison-mosaic}. We show each of the twelve simulations at time equal to one precession period as well as at t = 96 Myr for comparison. The labels at the bottom of each panel describe which simulation is shown and the simulation time represented. The simulations in the left two columns have a precession period of 3.2Myr, the 3rd and 4th column have a precession period of 31.7 Myr, and the 5th column has a precession period of 95.1 Myr. When comparing the jets all at the same dynamical stage (after 1 precession period) the reader should examine columns 1, 3, and 5 together. When comparing at the same time, the reader should instead compare columns 2, 4, and 5.

As Figure \ref{fig:comparison-mosaic} demonstrates, these systems evolve quite differently in terms of their dynamics. The jets with large precession angles do not extend as far in any one direction and are instead spread over a wider area, with cocoons of larger lateral extent. Similarly, jets with shorter precession periods are also more centrally condensed as they are unable to deposit momentum in a sustained direction long enough to substantially push back the denser ICM plasma. 

We can quantify the changes in dynamics using the ratio $\ell_b/(v_h \Delta t)$, as described in section \ref{subsec:propgation}. With our chosen jet injection parameters and equation \ref{eq:momentum2}, the propagation rate of the head of the jet through the ambient medium was 3.45 kpc/Myr. Given, this, we calculated the ratio $\ell_b /(v_h \Delta t)$ in Table \ref{tab:ratio}. By comparing the value of this ratio to the corresponding images in Figure \ref{fig:comparison-mosaic}, we can see that precessing jets with higher values of this ratio will be more tightly wound and generally have propagated less far from the source. Conversely, precessing jets with lower values of this ratio will be less tightly wound and propagate farther before curving and bending backward. While this ratio does not have a clear critical value that determines when a jet will break up from precession or display a clear s-shape morphology, it is still useful for comparing how different precession parameters lead to differing morphology. In particular, the dependencies on $\tau$ and $\psi$ are $\ell_b/(v_h \Delta t) \propto (\sin{\psi}/\tau)^{1/3}$.

Taking this ratio, as well as the general trends with precession period from Figure \ref{fig:comparison-mosaic}, it is clear that as the precession period decreases, the jet cocoons are more condensed and the jets more dramatically wound up. In simulations not presented here with shorter periods, the jet began to break up in our simulations. These jets may even have trouble breaking out of the interstellar medium of the host galaxy, which we do not include in these simulations.

Material surrounding the jet source is a combination of jet backflow during the initial stages of jet launching as well as material that has bled off from the jet as it precesses around. This material at late times radiatively steepens to $\alpha \approx -1.5$, while fresh material injected by the precessing jet remains spectrally flat. In the cases with large precession angle, there is more significant mixing of the older and newer jet material, as the cocoons are smaller and the jet injection covers a larger solid angle during their precession.

Examining the 5th column of Figure \ref{fig:comparison-mosaic}, with precession period P $\sim$ 95 Myr, we can see that as the precession period begins to approach the cooling timescale for the CRe, the surface brightness of the material away from the current jet direction drops off. If the period becomes too long, the CRe will age enough that they may no longer be detectable. This would leave the observer with little or no evidence of precession, and we would undercount precessing sources on the long period end of the distribution. However, only the longest lived AGN will reach such sustained jet ages, with most having lifetimes $> 100$ Myr \citep[e.g.,][]{Turner15}. If the jet precession has such a long period but isn't active for a large enough fraction of the precession period, the source will likely be indistinguishable from a non-precessing source. Because of this, radiative cooling may not be a major constraint on observing precessing radio jets.

\begin{table}
\begin{mdframed}[linewidth=0pt]
\centering
\begin{tabular}{c|c|c|c|c}
Simulation  & $\Delta t$ & $v_h \Delta t$ & $\ell_b$ & \multirow{2}{*}{$\frac{\ell_b}{v_h \Delta t}$} \\
Name  & (Myr) & (kpc) & (kpc) & \hfill \\\hline
\textbf{P3A10}  & 2.93 & 10.1 & 19.5 & 1.9\\ 
\textbf{P3A20}  & 1.5 & 5.2 & 12.4 & 2.38\\ 
\textbf{P3A30}  & 1.02 & 3.52 & 9.6 & 2.73\\ 
\textbf{P3A45} & 0.72 & 2.5 & 7.6 & 3.04\\ 
\textbf{P32A10}  & 29.1 & 100.4 & 89.7 & 0.89\\ 
\textbf{P32A20}  & 14.8 & 51.1 & 57.1 & 1.12\\ 
\textbf{P32A30}  & 10.1 & 34.9 & 44.3 & 1.27\\ 
\textbf{P32A45}  & 7.1 & 24.5 & 35.2 & 1.44\\ 
\textbf{P95A10}  & 87.2 & 300.8 & 186.6 & 0.62\\ 
\textbf{P95A20}  & 44.3 & 152.8 & 118.8 & 0.78\\ 
\textbf{P95A30}  & 30.3 & 104.5 & 92.2 & 0.88\\ 
\textbf{P95A45}  & 21.4 & 73.8 & 73.2 & 0.99\\ 
    \end{tabular}
    \caption{Values of jet bending analysis variables from section \ref{sec:precessioncartoon}. $\Delta t$ comes from Equation \ref{eq:dt}, $v_h$ from Equation \ref{eq:momentum2}, and $\ell_b$ refers to the definition in Equation \ref{eq:bending2}.}
    \label{tab:ratio}
    \end{mdframed}
\end{table}

\subsection{Effects of Viewing Angle}

\begin{figure*}
    \centering
    \includegraphics[width=0.85\textwidth]{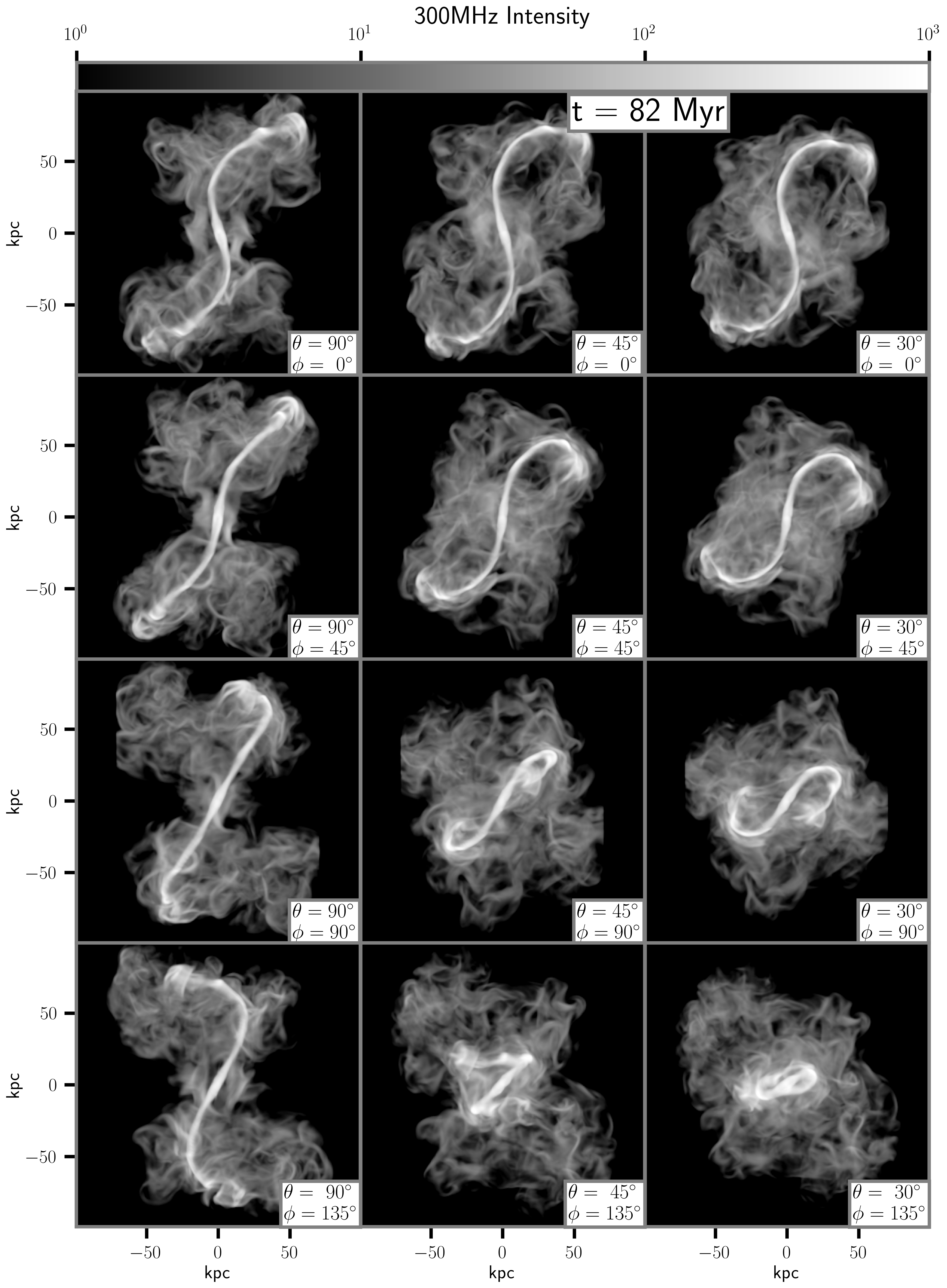}
    \caption{
    Radio intensity at 300MHz for the P32A30 simulation at t = 82 Myr from a variety of viewing angles. Radio images are convolved to 2.355'' FWHM resolution. Viewing angles are listed in the inset for each panel. For ease of visualization, the polar viewing angles are included in the diagram Figure \ref{fig:precession}. The polar angle is varied by column while the azimuthal angle is varied by row. In the online journal version, an animated figure is available (8 seconds, spanning 100 Myr simulation time) that follows the evolution of the source from each viewing angle simultaneously. Also, find the movie \href{https://www.youtube.com/watch?v=13_at_8ulF8&list=PLNl1xrpDmrjb0pcTUwSrTcgmpZrSYx81I&index=6}{here.}}
    \label{fig:jpr03-angles}
\end{figure*}

Within a single snapshot of a precessing jet system, its radio morphology may appear significantly different depending on the viewing angle. Figure \ref{fig:jpr03-angles} explicitly shows this for the P32A30 simulation. Each panel within this figure shows the P32A30 simulation at the same time (82 Myr), but from 12 perspectives. In the online version, this Figure is animated to show the evolution of this simulation in each of these 12 viewing angles simultaneously. At any given frame in the animation, one can see how the morphology changes with viewing angle. The differences are so stark that there are frames in which the radio source would very likely be classified in different ways, and lead to a misunderstanding of the physics of the source's evolution. For example, in the frame at $t = 82$ Myr, the first six frames (counting from left to right, top to bottom) and the 9$^th$ frame show signs of precession with curved jets. In the 7$^{th}$ and $10^{th}$ frames, the jet appears mostly straight, with some deflection at the farthest points. In the 8$^{th}$ and frame 11$^{th}$ frame, the jets appear to have been disrupted or sharply bent. An observer might misinterpret these sharp bends as signs of some environmental dynamics acting on the jets. Lastly, the radio source in the 12$^{th}$ would likely be unclassifiable. By providing the animation of the evolution of this source from many angles, we hope that radio observers can compare these images to detected radio galaxies to help identify possible precessing sources. We also hope to display the full range of complex morphologies that precessing jets may take on.

\subsection{Jet Reorientation Events}

\begin{figure*}
    \centering
    \includegraphics[width=0.7\textwidth]{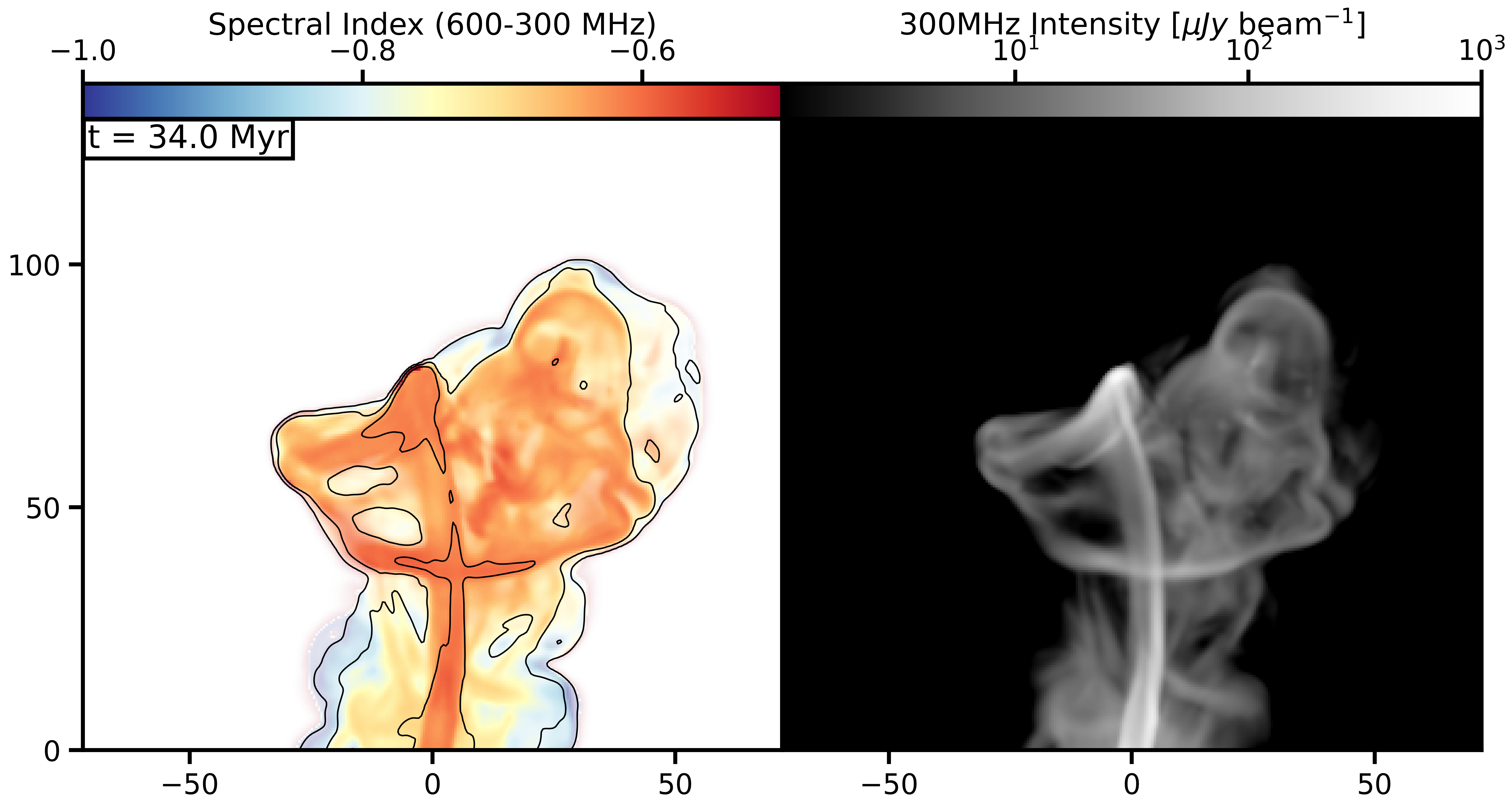}
    \caption{Left: Radio spectral index (600-300MHz, 
    weighted by 300MHz intensity with 300 MHz radio brightness contours (levels = [1, 10, 100]$\mu$Jy beam$^{-1}$) for the P32A30 simulation at t = 34 Myr. Right: 300MHz radio brightness at the same time. Radio images are convolved to 2.355'' FWHM resolution. At this time, the jet is undergoing a ``jet reorientation event'' in which the jet trajectory changes by more than 90$\degree$ over about 2.5 Myr. This can be seen here as the very sharp bend at the tip of the jet head at coordinates (x,y) $\approx$ (-5, 80) kpc. In the online journal version, an animated figure (5 seconds) shows the full duration of the reorientation event from t = 30 Myr until t = 38 Myr. Also, find the movie \href{https://www.youtube.com/watch?v=WGelAjxoFwQ&list=PLNl1xrpDmrjb0pcTUwSrTcgmpZrSYx81I&index=7}{here.}}
    \label{fig:reorient}
\end{figure*}

In each simulation, the precessing jets would sometimes undergo a dramatic and sudden change in their propagation direction. This ``reorientation event'' occurred roughly after one precession period had elapsed in the simulation. As the jet precesses initially, it bends strongly due to ram pressure from its interaction with the ICM. However, after one precession period has elapsed, the jet encounters a region of space in which it has previously deposited low density plasma. As the jet encounters this density discontinuity, the velocity of the head of the jet suddenly increases, as can be inferred from Equation \ref{eq:momentum2}. Additionally, the ram pressure that leads to the bending of the jet is greatly reduced, leading to an increase in the bending length (Equation \ref{eq:bending2}). Both of these effects lead to the jet becoming straighter and propagating to further distances before it then begins to interact with the higher density ICM when it reaches the outskirts of the low density cavity. This process is illustrated in Figure \ref{fig:reorient}, which shows the beginning of the reorientation event, in which the jet trajectory suddenly changes by $>90\degree$.
This type of sudden change in jet propagation direction is often attributed to either changes in the jet properties or duty cycle, or to external cluster dynamics such as shocks or other waves traveling through the medium. However, none of these effects are included in our simulations. These jet reorientation events can come about from jet precession alone, and the self interaction of the jet with its own previous activity. 

\subsection{Odd Radio Circles (ORCs)}

\begin{figure*}
    \centering
    \includegraphics[width=0.7\textwidth]{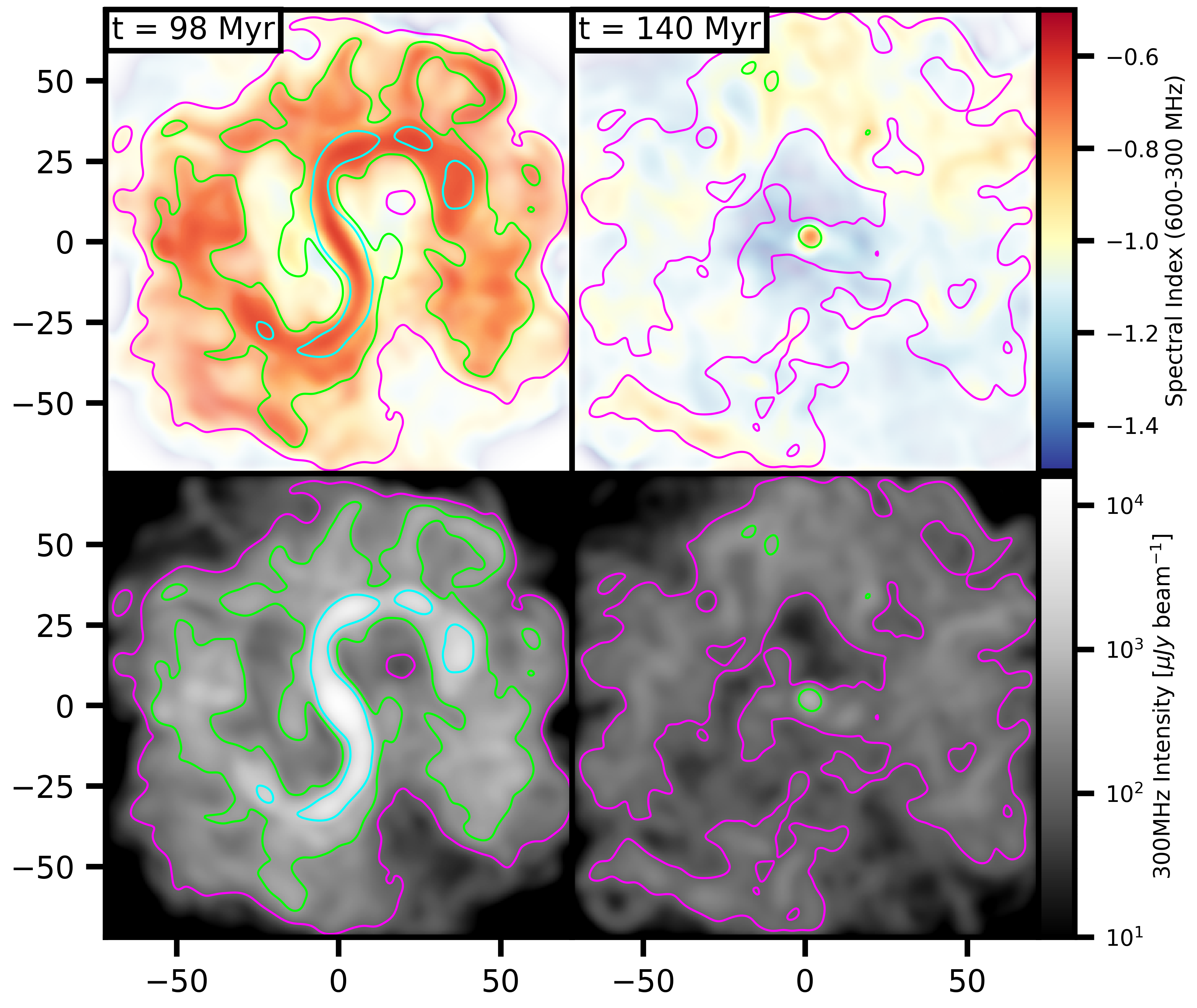}
    \caption{Top panels: Radio spectral index (600-300MHz, cut at 100$\mu$Jy beam$^-1$ at 300MHz) with 300 MHz radio brightness contours (levels = [100, 400, 1600]$\mu$Jy beam$^{-1}$) for the P32A30 simulation at 98 Myr (left) and 140 Myr (right) both viewed from above the jet precession axis ($\theta = 0 \degree$). Bottom panels: 300MHz radio brightness at the same times. Radio images are convolved to 11.775'' FWHM resolution. The jet turns off at 100 Myr and the plasma is allowed to radiatively cool. In the online journal version, a two-panel animated figure (10 seconds) follows the evolution of the source from the beginning of the simulation until t = 160 Myr. Also, find the movie \href{https://www.youtube.com/watch?v=zc5P2naNQa0&list=PLNl1xrpDmrjb0pcTUwSrTcgmpZrSYx81I&index=5}{here.}}
    \label{fig:jpr03-topdown}
\end{figure*}

Odd Radio Circles (ORCs) are recent and mysterious radio sources. They are circles of steep spectrum, diffuse radio emission. Some (but not all) ORCs have a detected galaxy near their center. There have been a variety of possible physical explanations as to their origin, including a spherical shock from a starburst wind, a supermassive black hole merger, or radio galaxy lobes seen end-on. Our simulations may lend some support to this last suggestion. Figure \ref{fig:jpr03-topdown} shows the P3A30 simulation along a line of sight parallel to the precession axis at two times. The jets turn off just after the time in the left panel. The right panel shows the radio emission about 40 Myr later. The radio morphology is similar to that of ORCs. In the online version of this article, an animated version of this figure is available, which shows the evolution up from 0-160 Myr, with the jet powering off around t=100 Myr. Just after the jet turns off, there is a full ring of emission visible. However, the radio spectrum is still relatively flat with $\alpha \approx -0.8$. The radio ring ages in place, without expanding further, as it is in pressure balance with its surroundings and the jets are no longer injecting momentum into the region. As the plasma radiatively ages and spectrally steepens, the ring becomes broken. There is a period of time in which the ring is still mostly intact and visible and it approaches the spectral index observed for some ORCs $\alpha 
\approx -1.2$ \citep{Norris22}. 

If this mechanism for explaining some ORCs holds, then it may be interesting to reexamine existing observations of S-shaped RGs and burn out the images looking for regions of low level emission surrounding the visible structures as a possible analog to the existing ORCs. This low level emission may not appear as circular from viewing angles misaligned from the precession axis, but it may represent the same material. Additionally, searches for AGN counterparts in galaxies near the centers of ORCs or VLBI observations looking for any current small scale jets could help determine if such a formation mechanism is possible.

\section{Summary}
\label{sec:Summary}

We have presented a series of simulations that look into the properties of precessing radio jets. These jets take on a wide variety of morphologies, depending on the properties of the precession itself as well as on the angle at which the source is viewed. In particular, the effects of viewing angle can greatly affect the classification of these sources. In many viewing angles, precession is not an obvious mechanism, as the jet may appear straight in projection or with very complex and messy morphology. 

Additionally, we observed some interesting and unexpected dynamics in our simulations, including self induced ``reorientation events," in which the jet trajectory quickly changed as the jet encounters a region filled with previous jet material of lower density than the ambient medium. This led to sharp turns in the jet which could be misinterpreted as due to external dynamics affecting the jet.

Lastly, we point out the similarities between ORCs and the remnants of a precessing jet seen end--on. This could be a possible explanation for some ORCs. Followup observations of existing ORCs looking AGN counterparts could help determine if this is likely. 

\acknowledgements

C.N. acknowledges funding through the NSF grant AST-1907850 as well as from Los Alamos National Laboratory through the LDRD program and NASA programs through the Astrophysical Theory Program. J.B. and T.M.N. were supported by NSF grant AST-1907850. We thank L. Rudnick, T. Jones, and P. C. Fragile for useful discussions.  The simulations presented here were run and analyzed at the College of Charleston on their high-performance Linux cluster (\url{https://hpc.cofc.edu}) as well as utilizing the Los Alamos National Laboratory Institutional Computing Program. 

\bibliography{Precession.bib}

\end{document}